\documentclass[letterpaper]{article}

\usepackage{aaai}
\usepackage{times}
\usepackage{helvet}
\usepackage{courier}
\usepackage{multirow}

\usepackage[table,xcdraw]{xcolor}

\usepackage{graphicx} 
\graphicspath{ {images/} }

\begin{document}
%
\title{Understanding Cognitive Depletion in Novice NMR Analysts}
\author{Lyndsey Franklin, Kyungsik Han, Zhuanyi Huang, Dustin Arendt, Nathan Hodas\\
Pacific Northwest National Laboratory\\
902 Battelle Blvd\\Richland, WA 99354
}

\maketitle

\begin{abstract}
\begin{quote}
We present the results of a user study with novice NMR analysts (N=19) involving a gamified simulation of the NMR analysis process. Participants solved randomly generated spectrum puzzles for up to three hours. We used eye tracking, event logging, and observations to record symptoms of cognitive depletion while participants worked. Analysis of results indicate that we can detect both signs of learning and signs of cognitive depletion in participants over the course of the three hours. Participants' break strategies did not predict or reflect game scores, but certain symptoms appear predictive of breaks.
\end{quote}
\end{abstract}

\section{Introduction}
\noindent Evidence suggests that cognitive performance decreases following periods of sustained mental effort and that taking breaks helps knowledge workers of all types from paid-per-task Mechanical Turk workers to full-time analysts of every domain \cite{ariga2011brief,borghini2014measuring,lim2010imaging,lottridge2015effects,rzeszotarski2013inserting,warm2008vigilance}. For instance, \cite{rzeszotarski2013inserting} asserts that workers are refreshed after breaks and that workers who take breaks perform more work compared to those who do not. It has also been observed that participant's resting baseline behavior predicts subsequent performance declines \cite{lim2010imaging}. It may be possible to improve a worker’s overall performance by improving his/her baseline between periods of task activity. But there is competition between the production rate of work from knowledge worker and the quality of their work. Highly motivated knowledge workers’ performance may decrease in quality if they continue to work for long periods of time without some sort of break. Knowledge workers focused on completion of tasks may not be able to accurately monitor their own fatigue level and may miss subtle cues indicating that they would benefit from some sort of break. 

We explore the notion of cognitive depletion where knowledge workers’ cognitive resources are depleted by continued work and their ability to accurately complete their task or provide meaningful analytical feedback becomes compromised. Cognitive depletion extends beyond simple sleep-deprivation related fatigue, and experts are susceptible to cognitive depletion because they may not feel (or admit to) the early signs of mental fatigue we often associate with exhaustion. It is important then to have a mechanism for identifying impending cognitive depletion  without the direct input of an individual to describe their current levels of fatigue and depletion. This mechanism must be able to reason about a person's state based on overt, observable symptoms in an unbiased way if it is to benefit its users.  Identifying such observable symptoms remains a challenge because fatigue, in general, is not well understood.

In this work, we present the results of a user study designed to evaluate a set of potential symptoms of cognitive depletion from a total of 19 participants. We use a browser-based game to simulate the analysis of Nuclear Magnetic Resonance (NMR) spectrum and asked participants to play this game for up to 3 hours. During this session researchers observed the participants for candidate symptoms of cognitive depletion while collecting game event logs and eye tracking data. We use the results of this study to draw inferences about how cognitive depletion manifests when people are asked to perform demanding problem-solving tasks for long durations of time. Our work provides early insights into cognitive depletion and addresses the future research implications of the findings. The following are the contributions of this paper:

\begin{itemize}
    \item We summarize fourteen symptoms observed over the course of three hours of the user study (19 participants).
    \item We describe characteristics of each symptom and its relationship with others through the triangulation of different datasets (e.g., symptom observations, event logging, and eye-tracking) as well as individual- and group-level analyses.
\end{itemize}

\section{Related Works}
Our work  understanding cognitive depletion builds on research in the related fields of attention, interruption and task resumption, multitasking, error and mistake frameworks, quality control, working memory, and cognitive overload.

\subsection{Attention, Interruptions, and Task Resumption}
Research into attention processes often attempts to measure and predict how long individuals can attend to a specific task. One important concept from attention research is the often observed vigilance decrement: after periods of sustained effort on a vigilance task, individuals’ begin to miss cues critical to their task or workflow~\cite{anderson2015polymorphic,ariga2011brief,pattyn2008psychophysiological,warm2008vigilance}. Task-unrelated thoughts (TUTs) are another important phenomenon from attentional research. Individuals required to focus on a task for long periods of time often report self-distracting thoughts that are unrelated to the task at hand~\cite{adler2013self,forster2009harnessing}. Suppressing TUTs is a current topic of research but for our purposes, we view TUTs as a potential symptom of cognitive depletion. Interruption and task resumption research seeks to understand workflows and predict the optimal time to interrupt an individual so that the interruption is as minimally detrimental as possible (see our references for just a few examples). Such understanding is important to the study of cognitive depletion because any coping or mitigation strategies must also be minimally intrusive to workflows.

\subsection{Quality Control, Error, and Mistake Frameworks}
The study of Mechanical Turk-style economies has provided a wealth of techniques for assessing the quality of worker contributions and for detecting workers who are abusing task structures for their own gain~\cite{cheng2015measuring,rzeszotarski2011instrumenting,salvucci2009toward}. These techniques are beneficial to the study of cognitive depletion as they provide potential metrics that can be automatically collected without interrupting workers as well as providing insight into the working patterns of large groups of individuals. It also provides an easily accessible real-world example of an ideal scenario for cognitive depletion research: an economy based on constant completion of micro-tasks where workers are motivated to work beyond the point where their cognitively depleted state begins affecting the quality of their output.

A tremendous amount of work has been done to understand and categorize human error in a variety of contexts. This work is particularly important in fields such as air traffic control where human error endangers hundreds of lives~\cite{kontogiannis2009proactive,zuger2015interruptibility}. Our work and model presented here do not attempt to replicate the in depth error frameworks completed by others. Rather, we use them as a source of observations and discrete observable phenomena which may be symptoms of cognitive depletion.

\subsection{Working Memory and Cognitive Overload}
Research into working memory and cognitive overload are the closest parallels to cognitive depletion that we have found and numerous works have provided the foundation for this current work. From these works important concepts such as mental and cognitive fatigue have arisen~\cite{galy2012relationship,guastello2012catastrophe,van2003mental}. Such research makes important distinctions between the effects of acute workload and extended engagement. Particularly, a sudden acute workload may cause a sort of `buckling stress' where an individual is unable to cope, while fatigue is the gradual loss of work capacity~\cite{guastello2012catastrophe}. \cite{van2003mental} defines mental fatigue as a change in psychophysiological state due to sustained performance. Both working memory and cognitive overload research study the immediate effects of task load in order to determine when a person is cognitively overloaded. We view cognitive overload as one mechanism which will, with extended engagement, result in a state of cognitive depletion. Highly overloaded persons will reach a state of cognitive depletion at a rate faster than those who are not cognitively overloaded. Mental fatigue as defined by \cite{van2003mental} is then a cue that an individual is cognitively depleted. One seminal work of this field is the frequently used NASA-TLX~\cite{hart1988development} which provides a six-axis inventory for self-assessing the cognitive load of a task. Our work builds on this foundation to assess signs indicating that a person may be in a state of cognitive depletion.

\section{Symptoms of Cognitive Depletion}
Building on the discussed related fields, we drafted a collection of potential symptoms of cognitive depletion based on the results of published experimental evaluations (see Table \ref{tab:summary_symptoms}). This gave us a starting collection of possible symptoms that we can group into three broad categories as follows.

\begin{table}[t!]
	\centering
	\footnotesize
	\begin{tabular}{p{1.5cm} | p{1cm} | p{4.5cm}}
		\hline
		\bf Category & \bf Code & \bf Symptom \\ \hline
		VR  &   D   & Distraction                       \\
		VR  &   I   & Inattention                       \\
		VR  &   IDT & Increased Decision Time           \\
		VR  &   TUT & Task Unrelated Thoughts           \\ \hline
		PED &   DCE & Data/Command Errors               \\
		PED &   INA & Increased Negative Affect         \\
		PED &   PE  & Physical Effects                  \\ \hline
		PJS &   C   & Confusion                         \\
		PJS &   ERE & Effort/Risk Over/Under Estimation \\
		PJS &   F   & Forgetting                        \\
		PJS &   LAP & Lack of Advancement/Progress      \\
		PJS &   STI & Strategy Inefficiency             \\
		PJS &   TA  & Task Abandonment                  \\
		PJS &   TR  & Task Rushing                      \\  \hline 
	\end{tabular}
	\caption{Symptoms and symptom categories under study. Categories are Vigilance and Reactionary Symptoms (VR), Physical and Emotional Deregulation (PED), and Personal Judgment Symptoms (PJS)}
	\label{tab:summary_symptoms}
\end{table}

\subsection{Vigilance and Reactionary Symptoms (VR)}
Vigilance and reactionary symptoms include phenomena such as habituation to status alerts, increased reaction time, distraction, and inattention. These symptoms are most likely to manifest in conditions require active monitoring of real-time events. For example, `vigilance decrement' is a well known phenomenon where reaction time and attentiveness to a  stimuli stagnates as function of time task~\cite{lim2010imaging,pattyn2008psychophysiological,rzeszotarski2011instrumenting,warm2008vigilance}. In this study, we focus on four potential symptoms: distraction (D), inattention (I), increased decision time (IDT), and task-unrelated thoughts (TUT).
\subsection{Physical and Emotional Deregulation (PED)}
Physical and motor symptoms include drowsiness, exhaustion, clumsiness, etc. as well as unintentional mistakes providing input, data, or keystrokes involving fine-motor skills. For example, a person may begin to fidget or change their seated posture rapidly or begin to click the wrong buttons on an interface~\cite{kontogiannis2009proactive,lim2010imaging,rzeszotarski2013inserting,gevins2003neurophysiological,adler2013self}. This category also includes emotional effects such as increased negative affect where a person may begin to complain about the difficulty of a task~\cite{adler2013self,bixler2013detecting,carver2003pleasure}. In this study, physical and emotional deregulation manifests as data/command errors (DCE), increased negative affect (INA), and physical effects (PE).
\subsection{Personal Judgment Symptoms (PJS)}
Personal judgment symptoms include occasions where a person is not accurately accounting for the state or their task and is confused or rushing tasks. It also includes disruptive workflows and the failure to form or adjust strategies. The distinguishing characteristic of symptoms in this category is that the person intends the actions resulting from an inaccurate judgement. If asked, the person would not say that he/she had made a mistake in the same sense that a person who misspelled a word might. In the latter case, the person knew how to spell the word but physically hit the wrong key. In the former, a person may intentionally use the word "then" when the grammatically correct word would be "than". Another symptom belonging to this category would be a demonstrated fixedness or rigidity in strategy towards a task even when it repeatedly fails~\cite{adler2013self,butler2015automatic,van2003mental}. Symptoms such as confusion (C), effort/risk under/over estimation (ERE), forgetting (F), task abandonment (TA), lack of advancement/progress (LAP), strategy inefficiency (STI), and task rushing (TR) belong to this category.

\section{Methodology}
Here we describe the environment and protocol we used to investigate cognitive depletion. Our study involved participants playing a puzzle-solving game for several hours while we collected data on overt symptoms of cognitive depletion.

\subsection{Apto Game Environment}
\begin{figure}[t]
\centering
\includegraphics[width=0.45\textwidth]{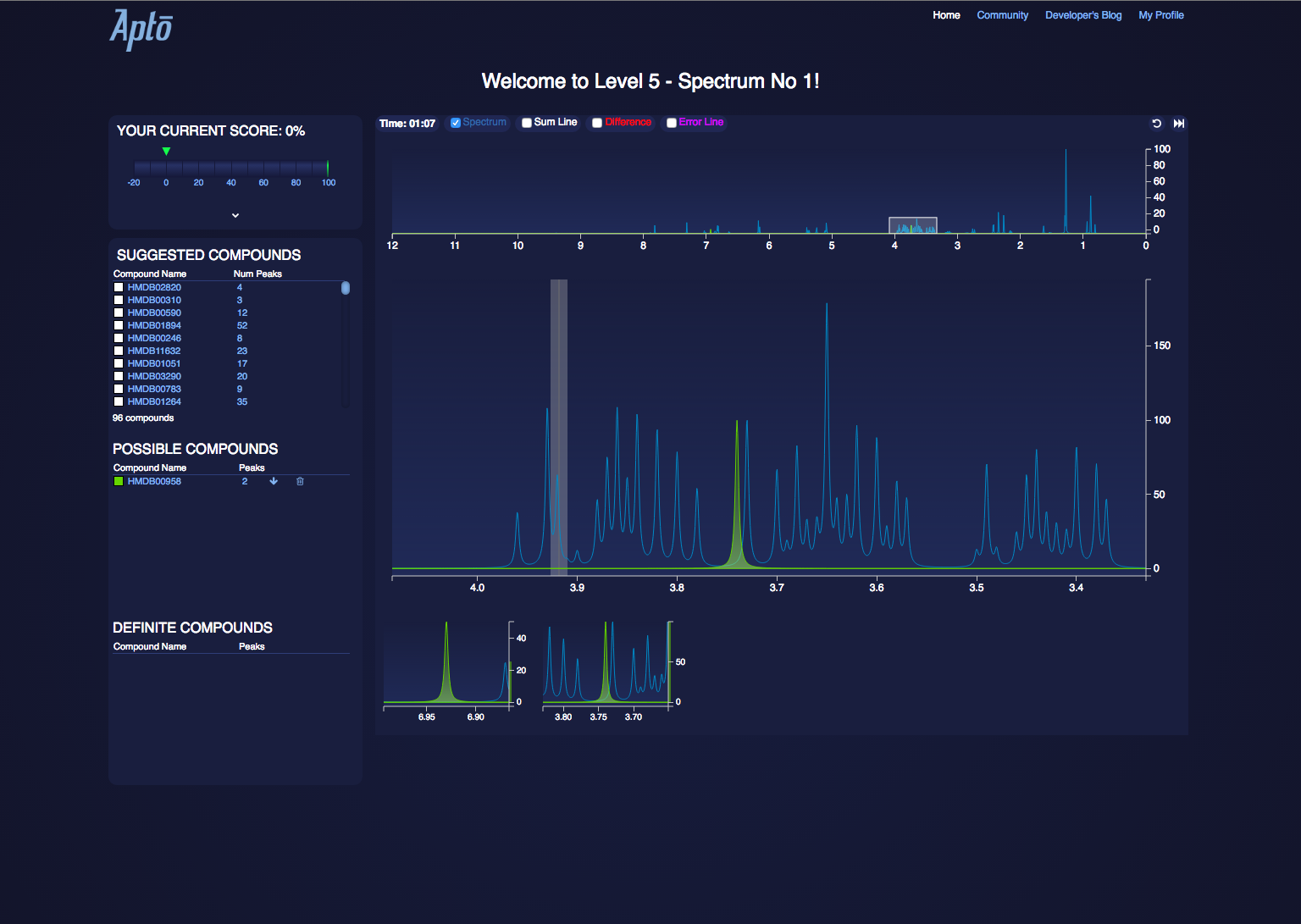}
\caption{The Apto Game Environment.}
\label{aptoss}
\end{figure}

Participants were recruited to play a web-game, Apto (see Figure \ref{aptoss}), which simulated the Nuclear Magnetic Resonance (NMR) spectrum analysis process and was designed to be similar to the professional software packages such as The Chenomx NMR suite~\footnote{http://www.chenomx.com/software/} used by experts to analyze spectrum results. Each spectrum was randomly generated and consisted of 1 to 16 compounds depending on the difficulty level. The score for each puzzle was calculated based on a formula that awarded more points per compound to higher levels. During the game, these scores were abstracted to a star system based on the game level: participants earned one bronze star for each completed level 1 puzzle, 2 bronze stars for level 2, 3 bronze stars for level 3, 1 silver star for level 4, and 1 gold star for each level 5 puzzle. The total score earned by a participant was available to him/her in this star form through a slide-down window. 

Participants solved puzzles by clicking peaks of the  spectrum to search for compounds with peaks near the search point. Participants selected candidate matching compounds from the search results of suggested compounds and then adjusted the fit of each compound to match as precisely as possible by clicking and dragging the compound's individual spectrum up or down relative to the puzzle spectrum. Compounds could have closely overlaping spectra, so three additional tools were available to help participants. A "sum line" showed the summation of the candidate compounds the participant currently had selected, so they could compare against the puzzle. A "difference line" drew a red line along the spectrum to highlight the difference between the sum line and the spectrum. An "error line" drew a bright-magenta line to draw attention to areas where the mean squared error between the sum and spectrum lines was great. These three tools operated independently and could be used in any combination by the participant. Participants had a "restart puzzle" button available to them which undid all their actions on a puzzle and returned it to its initial state. A "skip" button allowed participants to abandon the current puzzle in an unfinished state and be given a new puzzle to try. The restart and skip buttons did not affect the participant's score. When a participant completed a puzzle, they were given the option to change their level and could jump to any level from 1 - 5. Puzzles were considered solved when participants achieved 99\% of the available points. 


\subsection{Participants}
We recruited 19 participants (8 female) from our organization through flyers, emails, and meeting announcements. 9 of these participants wore glasses. Participants ranged in age and job description but did not have prior experience conducting NMR analysis. Participants were compensated for their time away from their regular jobs. In addition, we offered a small prize to be awarded to the participant with the highest session score at the end of the study.

\subsection{Protocol}
Our procedures were reviewed and approved by our IRB. Participants arrived and signed a consent form which explained what data would be collected and what their task for the session would be. Next, study facilitators attempted to set up the eye tracking device. Participants then completed an in-game tutorial which taught them how to query the system for suggested compounds and fit compounds into a spectrum to solve the puzzle. During this tutorial period, participants could ask for as much help as they needed from study facilitators to feel comfortable. When the tutorial finished, participants were given an additional tip-sheet which reminded them of how to use game controls and provided some basic strategies for solving complicated puzzles. Participants were then directed to play the game for the entire 3 hour session. Because we hypothesized that taking breaks related to cognitive depletion, we instructed them that they could take breaks whenever they chose, but we neither encouraged nor discouraged the users to take breaks. A study facilitator was present for the entire session and collected observational data by hand.

\subsection{Data Collection}
The game participants played automatically collected event information as the participants interacted with it. From this data, we have information about how many puzzles participants solved, how many compounds were in each puzzle (which refer to difficulty levels), and click data representing participant interactions with controls. We also collected timing information about how long it took for each puzzle to be solved through this game log. A Tobii Glasses 2 Eye Tracker~\footnote{http://www.tobiipro.com/product-listing/tobii-pro-glasses-2/} was used to collect eye movement, accelerometers, pupil dilation, and fixation data. We could not calibrate the eye tracker and collect eye-tracking data on 6 of the 9 participants who wore glasses. For these 6 participants, we rely on event log and observational notes to capture activity during the session. The eye tracker was turned off when participants took breaks so that they could leave to use the restroom and then re-calibrated upon their return from break. Study facilitators used a predetermined coding scheme to capture the symptoms observed during the session as well as general notes about the session such as errors encountered, break times, and any interrupting factors. To account for differences in timing devices and the reaction-time of study facilitators, the observed time of each symptom was recorded at the minute level.

After the study, for data reliability, four people (including two study facilitators and two researchers) looked into the logs and made sure that all the symptoms were all correctly and consistently annotated based on the symptom descriptions.


\section{Results}
\subsection{Study Engagement}

\begin{table}[t]
	\centering
	\footnotesize
	\begin{tabular}{p{0.8cm} | p{0.7cm} | p{0.7cm} | p{1cm} | p{0.8cm} | p{0.8cm}}
		\hline
		\bf Parti-cipant & \bf Avg. Difficulty & \bf \# Completions & \bf \# Skips & \bf Score & \bf Break \\ \hline
		P1  & 2.35 &  90 &  5 & 273K & Yes \\
		P2  & 3.48 &  23 &  0 & 431K & No \\
		P3  & 1.97 & 169 &  4 & 778K & Yes \\
		P4  & 3.63 &  35 &  3 & 578K & Yes \\ 
		P5  & 1.81 & 127 &  3 & 501K & Yes \\
		P6  & 2.76 &  95 & 10 & 1.1M & Yes \\
		P7  & 3.58 &  29 &  1 & 581K & Yes \\
		P8  & 3.60 &  45 &  0 & 767K & No \\
		P9  & 4.10 &  49 &  0 & 1.4M & Yes \\
		P10 & 2.30 &  62 & 10 & 160K & Yes \\ 
		P11 & 4.58 &  61 &  0 & 2.8M & No \\
		P12 & 2.55 &  78 &  2 & 331K & Yes \\
		P13 & 1.00 &  39 &  2 &  37K & Yes \\
		P14 & 3.79 &  22 &  1 & 425K & Yes \\
		P15 & 1.01 & 226 & 18 & 221K & Yes \\
		P16 & 3.05 &  41 &  2 & 555K & No \\
		P17 & 3.78 &  31 &  2 & 626K & Yes \\
		P18 & 4.42 &  28 &  0 & 799K & No \\
		P19 & 2.13 &  18 &  5 &  51K & Yes \\ \hline
	\end{tabular}
	\caption{Summary of task engagement by participants. Five participants did not take a break during the study. Participants who did not take a break completed more difficult tasks and less skipped than those who took a break (p$<$.05).}
	\label{tab:summary_participant}
\end{table}

A total of 19 participants engaged in the study differently with respect to average level of task difficulty, the number of trials and completions, the number of skips, total game score, and break (Table~\ref{tab:summary_participant}). Each participant showed different degrees of engagement but we found some correlations among the variables. Participants who worked at higher difficulty levels solved and skipped fewer puzzles ($p<$.05) but also spent more time on these difficult puzzles and so attempted fewer. We consider a break as variable and found that participants who did not take a break completed more difficult tasks, higher scores, and had less skips than those who took a break ($p<$.05). 

We were interested in which types of symptoms and how much of each appeared in general. In addition, as we found  different activity levels between the non-break and break groups, we decided to look into symptoms at the group level. \emph{We did not find any differences in symptoms between difficulty levels.} 

In the following subsections, we will first summarize 14 symptoms identified in the study and then report characteristics of the symptoms from all participants as a whole and at the group level. 

\subsection{Observed Symptoms}
Symptoms were collected through observations and cross-checked with inter-rater agreement, game event logs, and eye tracking data when possible (the eye tracker would not calibrate on 5 participants).  Symptoms in our study were operationalized as follows:


\begin{itemize}
    \item \textit{Distraction} (D, appeared 17 times) was recorded when participants showed signs of attending to stimuli that did not come from the Apto game environment. This included briefly re-orienting towards hallway noise, diverting attention to their cell phone if it rang, etc.
    \item \textit{Inattention} (I, N=30) was typically characterized as a participants lack of correction of errors: i.e. some participants picked poorly fitting compounds which caused them to lose points, however eye tracking data confirms they did not always look at their score to pay attention to the effects of their actions.
    \item \textit{Increased decision time} (IDT, N=73) and \textit{Lack of advancement/progress} (LAP) were both judged based on the time it took a participant to take an action and complete a puzzle relative to their performance during the rest of the session. Participants who took longer than usual to make a single decision were noted for IDT. Participants who took longer than usual to solve a puzzle or never attempted puzzles of higher levels were noted for LAP.
    \item \textit{Data/Command errors} (DCE, N=34) manifest as actions which were immediately undone and followed by a nearly identical action with preferred outcome (frequently paired with a audible ''Oops'').
    \item \textit{Increased Negative Affect} (INA, N=44) was recorded when participants verbal expressed emotional states such as frustration or impatience (''I'm still at 96\%? This is so frustrating!'').
    \item \textit{Physical effects} (PE, N=251) were the easiest symptoms for observers to confirm as they included events such as fidgeting and posture changes. The study protocol did not request or encourage a ''talk-aloud'' protocol so participants who chose to speak were often expressing confusion (''I don't know why that happened''), forgetfulness (''How do I scroll again?''), distraction(''The browser is running very slow''), effort/risk over/under estimation (''This is a tough one!''), negative affect (''This [puzzle] is taking forever''),or task-unrelated-thoughts (''I'm glad I participated today, this study is interesting!''). These verbalizations were not counted as PE.
    \item \textit{Confusion} (C, N=27) was typically noted through verbalizations such as "I don't understand" or "What did I just do?".
    \item \textit{Effort/Risk Over/Under Estimation} (ERE, N=6) manifested as level-increase actions followed by quick task abandonment and level decreases or verbalizations such as "This is harder than I thought!" and "Level 5 is much harder than it looks."
    \item \textit{Forgetting} (F, N=21) is indicated by participants asking study facilitators to repeat instructions covered in the tutorial or re-reading the tip sheet multiple times. In one case, a participant asked for scratch paper to use as a memory aid to write down the names of compounds she had looked at in solving a puzzle.
    \item \textit{Strategy inefficiency} (STI, N=53) typically manifest as a lack of help-line use: some participants manually searched for incomplete puzzle pieces instead of using the error or difference lines explicitly described in the tip-sheet.
    \item \textit{Task abandonment} (TA, N=62) was recorded as use of the game's 'skip puzzle' button.
    \item \textit{Task Rushing} (TR, N=33) is similar to IDT in that it represents a deviation from some baseline-level of participant actions. In the case of TR, this is a pace beyond what would be expected from experience and naturally improved skill. TR will likely manifest with DCE and I: participants rushing through a task make mistakes but do not correct them right away. TR also manifests as trial-and-error attempts to improve a participant's score on a puzzle (as opposed to carefully searching for a matching compound or using the error/difference/sum line tools).
    \item \textit{Task Unrelated Thoughts} (TUT, N=27) were noted when participants verbalized their thoughts which had no impact at all on their puzzle solving tasks such as whether or not the random spectrum matched real-world substances ("Is this hydrogen or something?") or even what researchers did in their free time ("You mush all meet up for drinks when this is over!")
\end{itemize}
Note that this list is not meant to be exhaustive. It is an overview of the symptoms we learned about from our study. Certain symptoms suggested by related work were not expected to be seen in this study. For instance, habituation occurs when people stop reacting to important information in the form of alerts \cite{anderson2015polymorphic}. The Apto game environment has no such alerts and therefor habituation (among other real-time-response-based symptoms) would not be detected in this study.

\begin{figure}[t]
\centering
\includegraphics[width=0.4\textwidth]{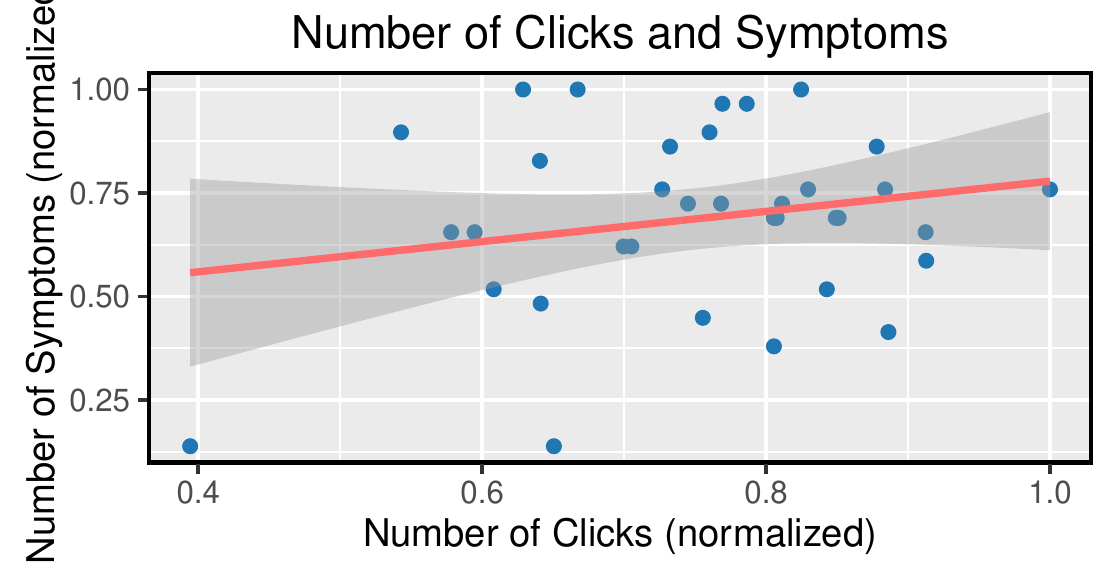}
\hspace*{.3cm}
\vspace*{-.3cm}
\caption{Correlation between the number of symptoms and clicks (normalized) in every one minute. Two variables showed a positive relationship ($r$=.23, $p<$.05).}
\label{Symptoms_clicks}
\end{figure}

\begin{table}[t]
	\centering
	\footnotesize
	\begin{tabular}{p{1.5cm} | p{1.2cm} | p{1.3cm} | p{1.2cm} | p{1.0cm}}
		\hline
		\bf Participant & \bf \# Symptoms & \bf Symptom Diversity & \bf \# Co-Symptoms & \bf Break \\ \hline
		P1  & 38 & 12 & 28 & Yes \\
		P2  & 36 & 11 & 28 &  No \\
		P3  & 39 & 12 & 21 & Yes \\
		P4  & 16 & 10 & 10 & Yes \\
		P5  & 34 & 12 & 15 & Yes \\
		P6  & 17 &  6 &  1 & Yes \\
		P7  &  9 &  9 &  1 & Yes \\
		P8  & 69 & 14 & 28 &  No \\
		P9  & 88 & 12 & 55 & Yes \\
		P10 & 39 & 11 & 36 & Yes \\
		P11 & 19 &  7 &  1 &  No \\
		P12 & 34 &  8 & 15 & Yes \\
		P13 & 41 &  7 & 21 & Yes \\
		P14 & 27 &  9 & 21 & Yes \\
		P15 & 52 & 13 & 15 & Yes \\
		P16 & 16 &  6 &  6 &  No \\
		P17 & 12 &  8 &  0 & Yes \\
		P18 & 45 &  9 & 15 &  No \\
		P19 & 75 & 14 & 91 & Yes \\ \hline
	\end{tabular}
	\caption{Summary of the number of symptoms and diversity by participants. The number of symptoms are strongly correlated with the diversity of symptoms ($r$=.75, $p<$.001) and the number of co-symptoms ($r$=.80, $p<$.0001).}
	\label{tab:summary_symptom_participant}
\end{table}

\begin{figure}[t]
\centering
\includegraphics[width=0.4\textwidth]{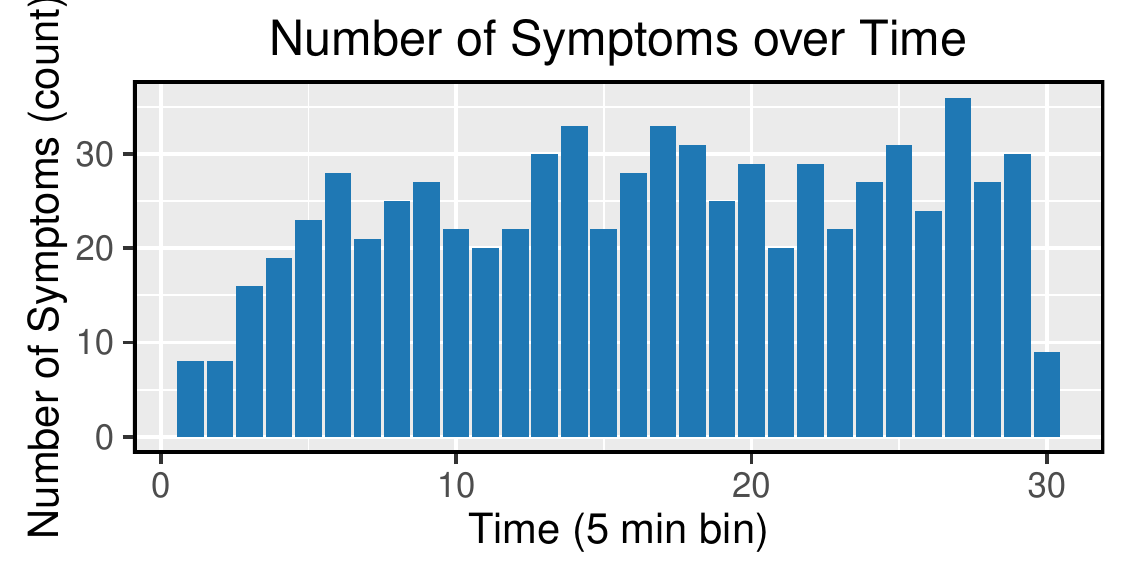}
\hspace*{.4cm}
\vspace*{.3cm}
\includegraphics[width=0.4\textwidth]{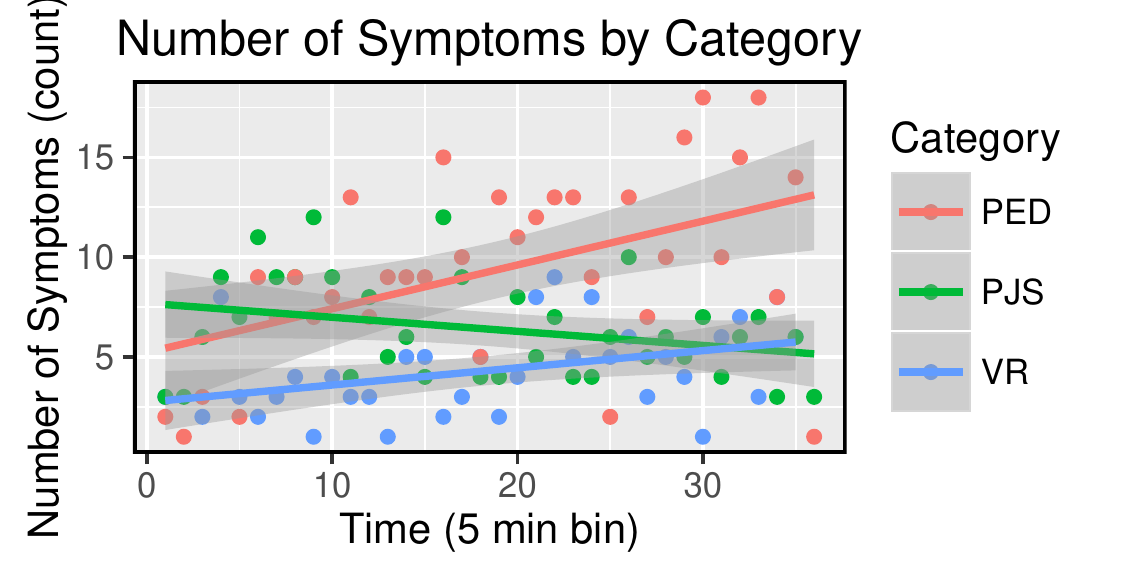}
\hspace*{.3cm}
\vspace*{-.3cm}
\caption{The number of symptoms during the study from all participants. Generally more symptoms appeared over time. For category, PED and VR increased ($p<$.05) while PJS decreased ($p<$.10) over time.}
\label{Symptoms_over_time}
\end{figure}

\subsubsection{Overall Summary}
From a total of 14 symptoms, physical effects (PE) were detected the most followed by increased decision time (IDT), increased task abandonment (ITA), etc. As shown in Table~\ref{tab:summary_symptom_participant}, more than 50\% of the participants (10) showed more than 10 symptom types. We also found that the number of symptoms was positively associated with the number of clicks, indicating that the participants tended to click more when they presented symptoms (Figure~\ref{Symptoms_clicks}).

\begin{figure}[h]
\centering
\includegraphics[width=0.45\textwidth]{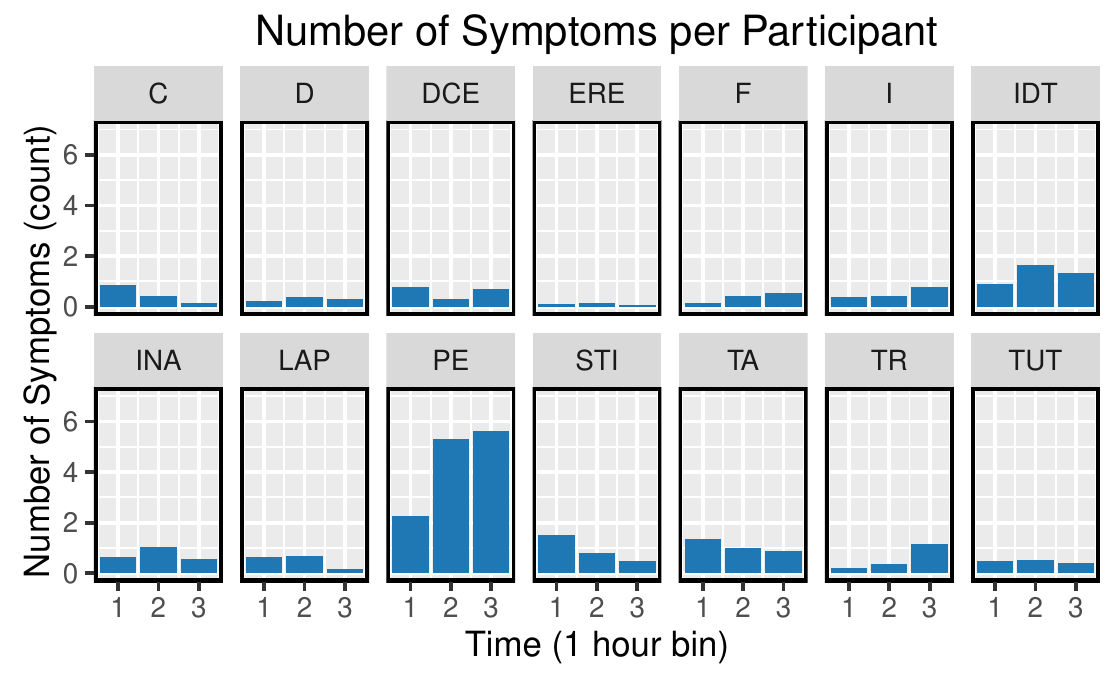}
\vspace*{-.3cm}
\caption{The number of symptoms per participant by type over time (1 hour bin). Physical effects (PE), forgetting (F), inattention (I), and task rush (TR) generally increased whereas strategy inefficiency (STI) and task abandonment (TA) decreased over time.} 
\label{symptoms_by_type_over_time_5}
\end{figure}

The number of symptoms generally increased over time (Figure~\ref{Symptoms_over_time}). We saw the significant drop at the end of the study. When we saw the results at the group level, PED and VR were found to be increased and PJS was decreased over time. As the participants progressed the study, it seems that they showed less task-related judgement symptoms but showed more physical and emotional symptoms. As shown in Figure~\ref{symptoms_by_type_over_time_5}, each symptom shows a different distribution over time. Linear regression results showed statistically significant increases in physical effects (PE) and inattention (I) and significant decreases in strategy inefficiency (STI) at $p<$.05. Given the length of the user study (3 hours), an increase in fatigue or decrease in concentration could contribute to PE and I. Learning effects could contribute to a decrease in STI because it is more pertinent to task efficiency. Although not statistically significant, TA generally decreased over time, also possibly due to a learning effect.

\begin{figure}[t]
\centering
\includegraphics[width=0.40\textwidth]{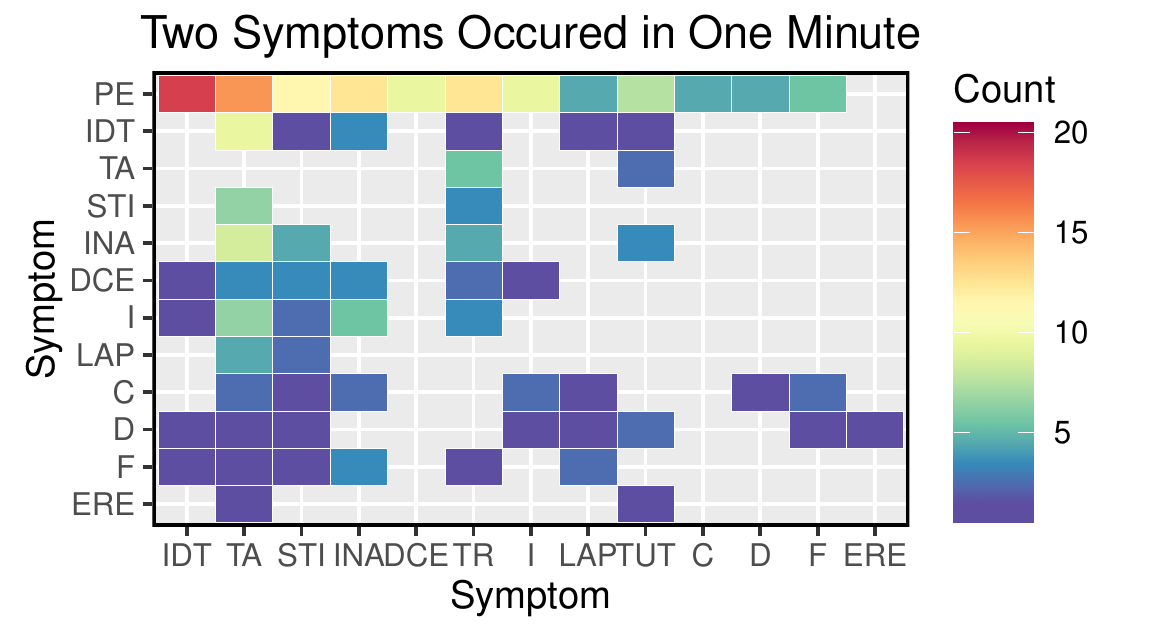}
\vspace*{-.3cm}
\caption{Two symptoms sequentially shown within one minute. A great number of symptoms occurred with others (64\%).}
\label{Symptoms_sequence}
\end{figure}

\begin{figure}[h]
\centering
\includegraphics[width=0.4\textwidth]{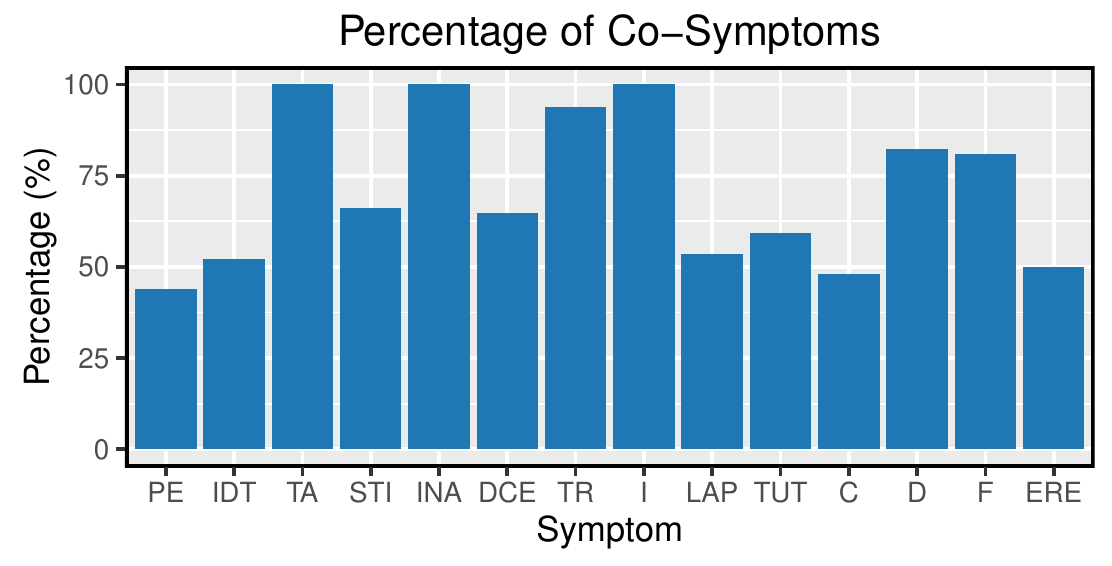}
\hspace*{.7cm}
\vspace*{-.3cm}
\caption{Percentages of symptoms occurred with other symptoms. Four symptoms (i.e., TA, INA, TR, and I) are most likely to show together with other symptoms (especially TA, INA, and I showed 100\% co-occurrence), whereas PE is relatively independent.}
\label{Symptoms_CoSymptoms}
\end{figure}

We also considered a possibility of two symptoms' appearing together or sequentially. Our observation logs indicated that there were lists of symptoms which together in the same one-minute window. We measured symptom pairs and their frequency. Out of 706 individual observations, a total of 456 symptoms in 62 groups of two symptoms (64\%) were identified. 

Figure~\ref{Symptoms_sequence} illustrates the results. As PE occurred the most during the study, it occurred together with many other symptoms. The symptom group with the top two symptoms showed the highest count. Figure~\ref{Symptoms_CoSymptoms} presents another aspect of co-occurrence. We measured the percentage of co-occurrence from the total number of occurrences for each symptom. Although the raw number of co-occurrences for PE was the largest, when it comes to percentage, it showed lower than 50\%. This indicates that PE also occurs independently and individually. This result applies to IDT (51\%) as well. In addition, TA, INA, TR, and I exhibited very high percentages (especially TA, INA, and I showed 100\%), indicating that these are the symptoms that could entail or be led by other symptoms.



\begin{figure}[t]
\centering
\includegraphics[width=0.45\textwidth]{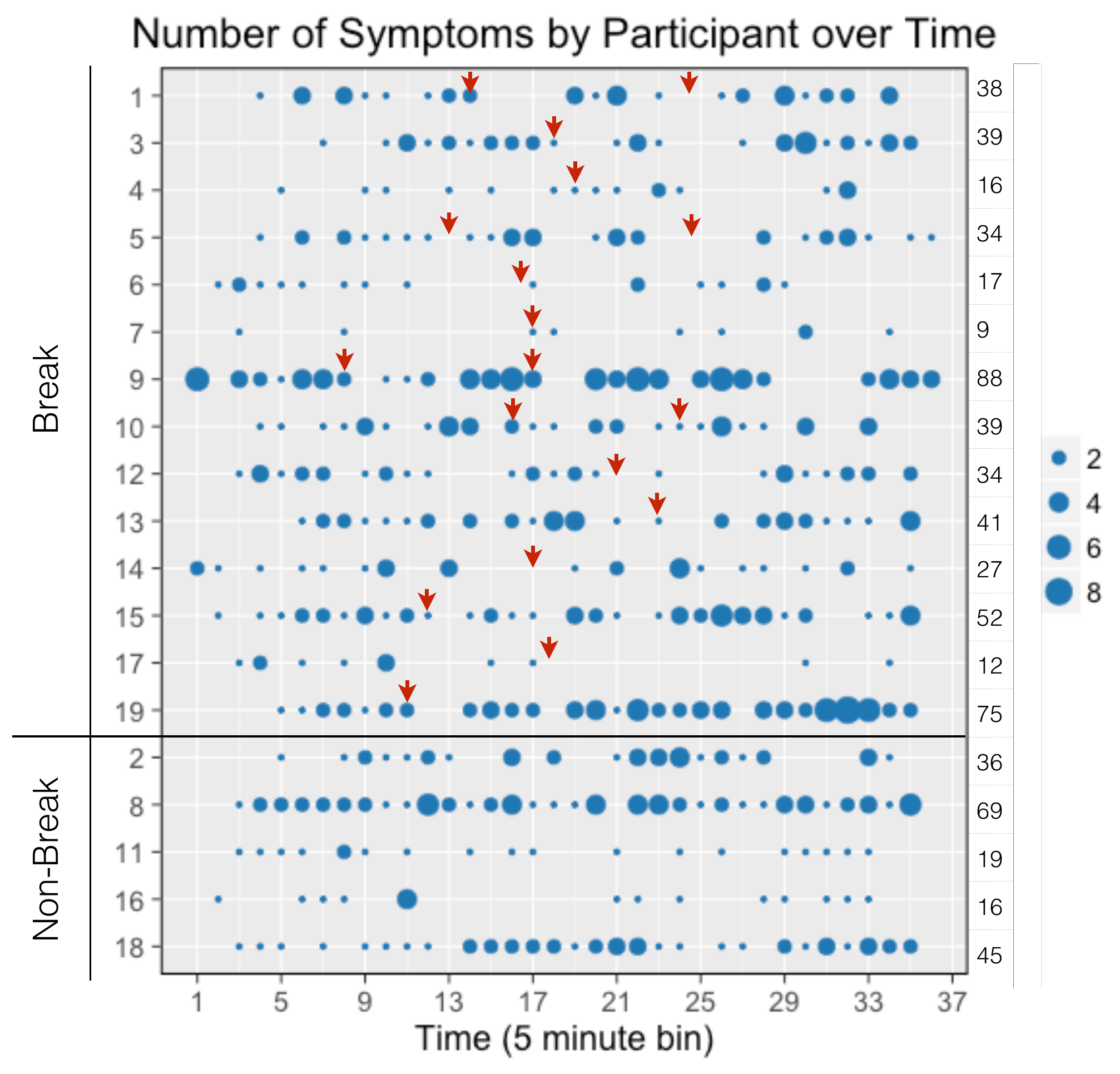}
\vspace*{-.3cm}
\caption{Percentage of symptoms in 10-minute bin for participants. Numbers on the right y-axis indicate total symptom counts. The first 14 participants were the ones who took break(s). Red arrow represents break. 6 participants (P1, P3, P9, P10, P15, P19) showed more than the average of symptoms at the same time slot or one time slot before the break. No symptom appeared during the same time slot after the break. The last 5 participants did not take a break.}
\label{symptoms_by_participant_over_time}
\end{figure}

\begin{figure}[t]
\centering
\includegraphics[width=0.42\textwidth]{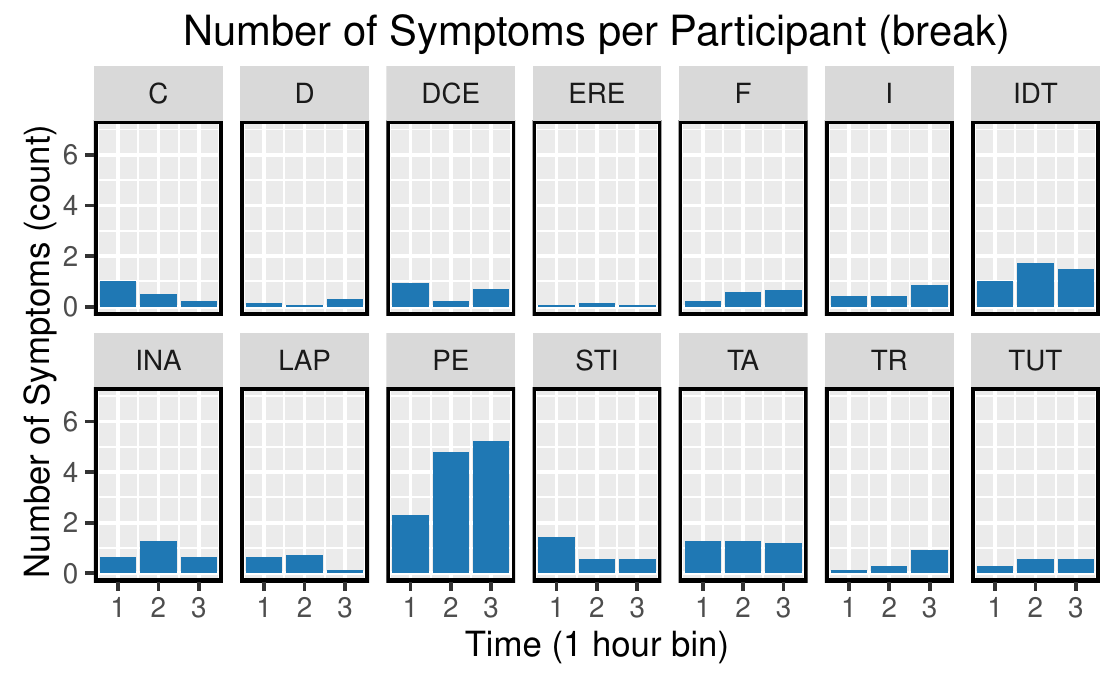}
\includegraphics[width=0.42\textwidth]{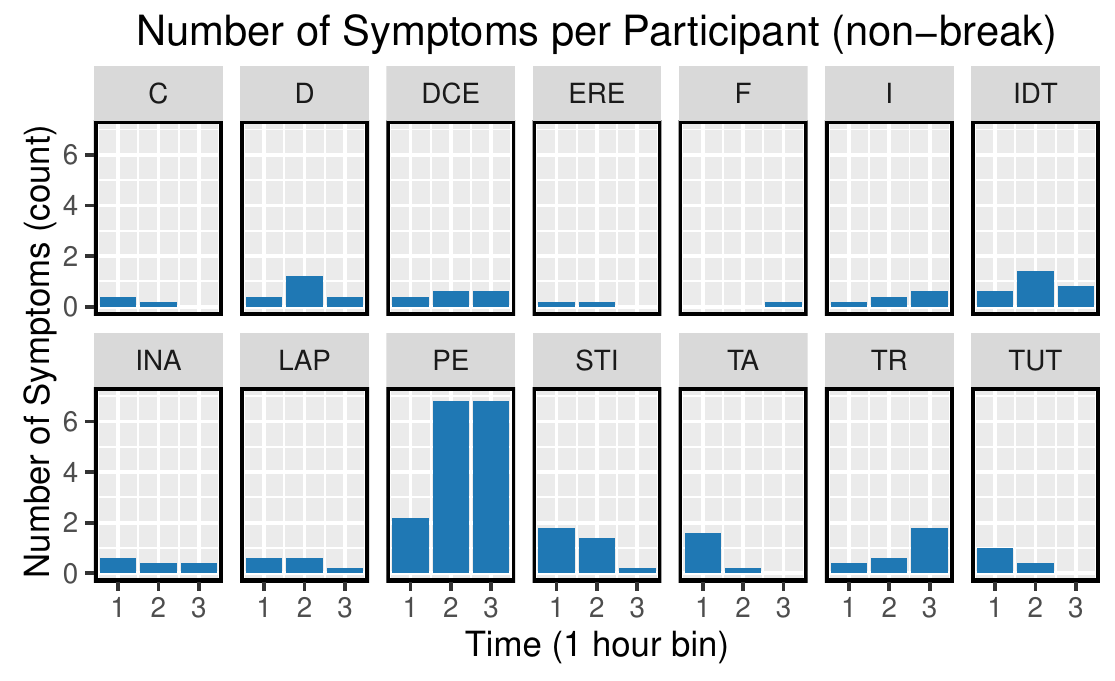}
\caption{Number of symptoms per participant (average) for the break (top) and non-break groups (bottom). Participants in the break group showed an overall increase in F, I, IDT, and PE, and decrease in C and STI. Participants in the non-break group showed an overall increase in I, PE, and TR, and decrease in STI and TA. Interestingly, the non-break group showed a great level of PE.} 
\label{Symptoms_by_type_over_time_5_break_non_break}
\end{figure}

\subsubsection{Symptoms by Participant and Break Group}

We now switch our focus to the results by an individual level and a group level. We had two user groups (i.e., break and non-break groups) for the analysis. Table~\ref{tab:summary_symptom_participant} summarizes the number of symptoms, symptom diversity, and co-symptoms which appeared in one minute (Figure~\ref{Symptoms_CoSymptoms}) by each participant. The number of symptoms is strongly correlated to symptom diversity ($r$=.75, $p<$.001) and the number of co-symptoms ($r$=.80, $p<$.0001). However there was no significant difference between break and non-break groups. 

Figure~\ref{symptoms_by_participant_over_time} illustrates the number of symptoms from each participant (clustered into either a break or non-break group) over time (5-minute bin). Red arrows represent breaks. Note that the break was found to be the last event of each time slot. We confirmed that no symptom appeared after the break for all participants. The results indicate that for quite a few participants (i.e., P1, P3, P9, P10, P15, P19) in the break group, the number of symptoms appeared at the same time of the break or one time slot before the break was above the average.









\begin{figure}[t]
\centering
\includegraphics[width=0.40\textwidth]{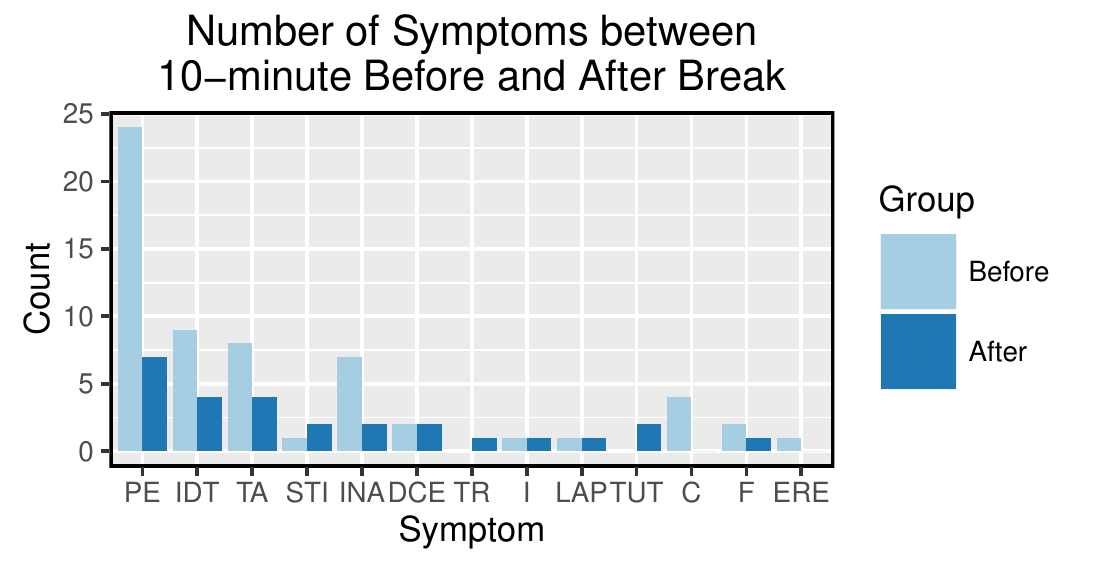}
\vspace*{-.3cm}
\caption{Difference in symptom percentage between 10 minute before and after break. Especially, physical effects (PE), increased decision time (IDT), task abandonment (TA), increased negative affect (INA), and confusion (C) appeared more before the break.}
\label{Symptom_diff_before_after_break}
\end{figure}

Figure~\ref{Symptoms_by_type_over_time_5_break_non_break} shows a variance of each symptom by the break and non-break groups. On the one hand, the break group showed the results that were quite consistent with the ones from all participants (see Figure~\ref{symptoms_by_type_over_time_5}). F, I, IDT, and PE generally increased while C and STI decreased. On the other hand, for the non-break group three symptoms (i.e., I, PE, and TR) increased and other three symptoms (i.e., STI, TA, and TUT) decreased. Especially for TA and TUT, two groups showed quite opposite results. It appears that the participants in the break group tended to be less focused than those in the non-break group. Another interesting finding is that the non-break group showed more PE than the break group.

We were interested in the effects of breaks and investigated the differences in the types and frequencies of symptoms during the 10 minutes before and the 10 minutes after breaks. Figure~\ref{Symptom_diff_before_after_break} showed several interesting insights. First, PE, C, and ERE appeared more before the break. PE appeared 17 times more before breaks, which indicates it could be a sign of one's intention to take a break. Second, for some symptoms, we found differences in frequency between their overall trends (see Figure~\ref{Symptoms_by_type_over_time_5_break_non_break}, top) and break periods. PE and IDT generally increased over time, but when it comes to the time around breaks, they appeared more before the break. TA, INA, and C appeared more before the break. Participants showed more physical effects (PE), increased decision time (IDT), task abandonment (TA), negative effects (INA), and confusion (C) before they decided to take a break. This indicates that these symptoms could be considered as possible signs of cognitive depletion and factors that could indicate that someone is in need of a break.

\begin{figure}[t]
\centering
\includegraphics[width=0.4\textwidth]{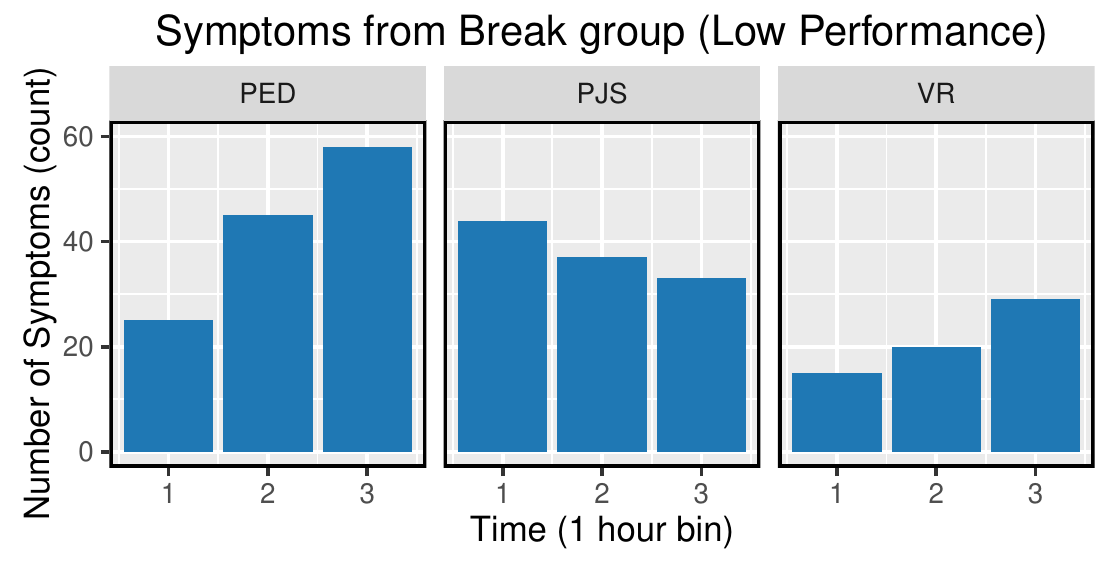}
\includegraphics[width=0.4\textwidth]{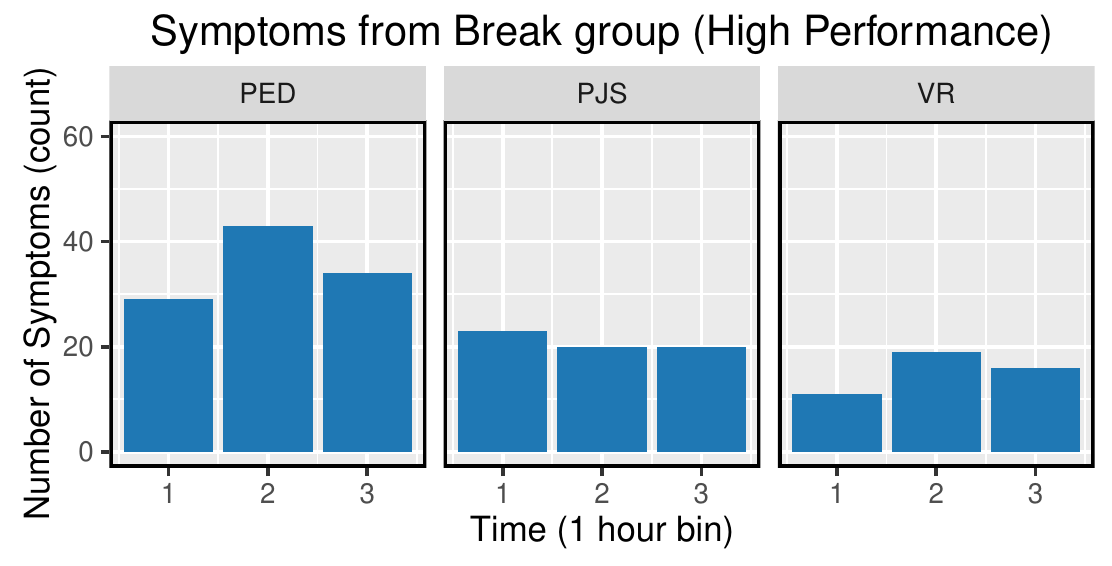}
\vspace*{-.3cm}
\caption{Difference in symptom groups between low- (top) and high (bottom) -performance break groups. The low group showed steady increases in PED and VR while the high group showed decreases from 2nd and 3rd time slots.}
\label{Symptoms_by_low_and_high_performance_group}
\end{figure}

Our last investigation in this section was looking into the participants in the break group. In the previous section, we showed generally lower performance by the break group than the non-break group. However we realized that some participants actually showed good performance and engagement (e.g., P6, P8). Thus, we measured differences in symptoms between two groups of participants based on their performance. We used the median of the scores to have the high and low groups (7 participants were assigned to each group). Figure~\ref{Symptoms_by_low_and_high_performance_group} shows the results. Both the low and high groups showed a decrease in PJS; however, showed different results in PED and VR. This low group still showed increases in PED and VR even they took breaks during the study. 

\subsubsection{Symptoms with Eye-Tracking Data}

\begin{figure}[t]
\centering
\includegraphics[width=0.40\textwidth]{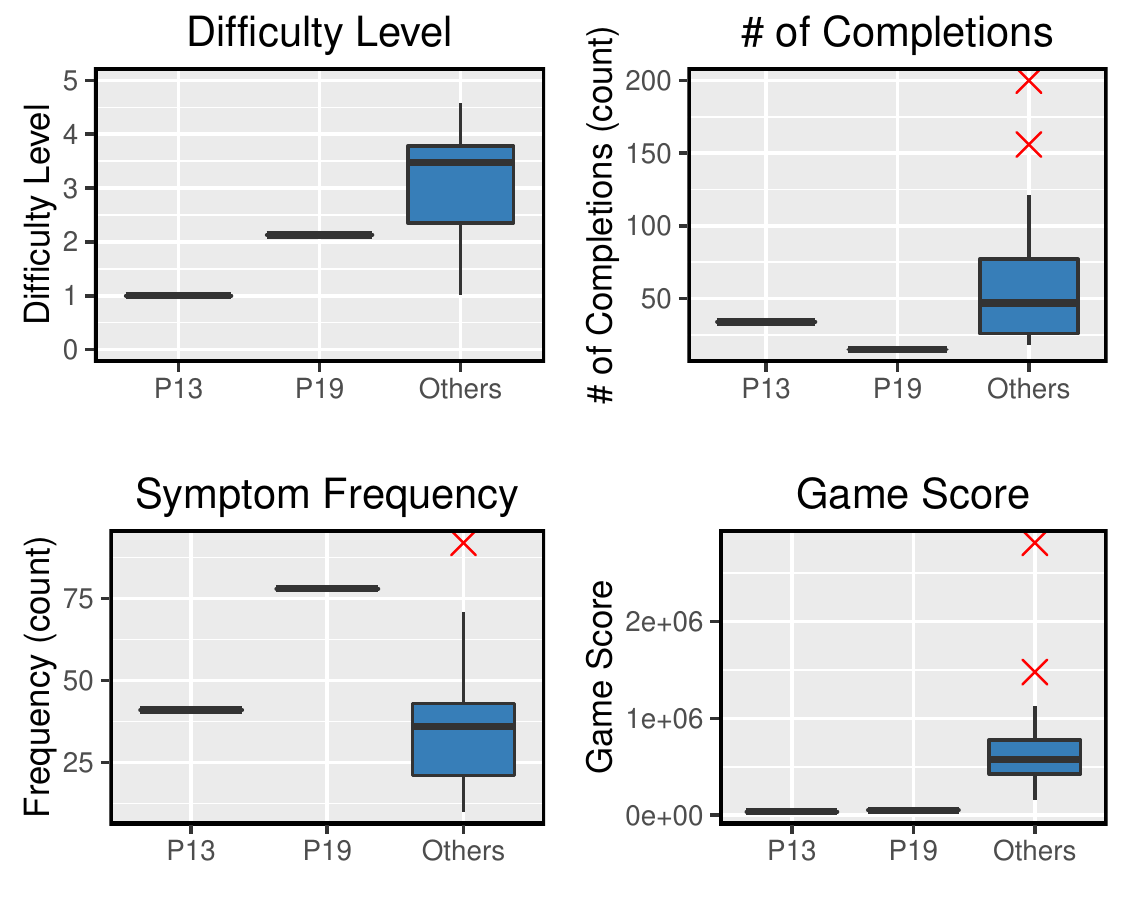}
\vspace*{-.3cm}
\caption{Difference in the level of activity and participation between P13 and P19 (who showed the positive relationship between pupil dilation and symptom frequency), and the rest of the participants. Both P13 and P19 worked primarily on low-level tasks, had significantly fewer task trials, more symptoms, and lower game scores. This clearly indicates the effect of cognitive loads and depletion on task performance. Red cross mark indicates an outlier.}
\label{P16_P23_and_Others}
\end{figure}

Our last investigation was to map the eye-tracking data to the observed symptoms. A pupil dilation response points to a level of emotion, arousal, stress, pain, or cognitive load of a human. In this regard, we were especially interested in the relationship between pupil dilation and the number of symptoms both by an individual level and a group level. For the group level, Spearman's correlation results (which is a nonparametric measure of the strength and direction of association) show that both break and non-break groups did not show any significant difference in pupil dilation. 

However, when looking into individual differences, we found that two participants from the break group showed significantly positive relationships (P13: $\rho$=0.28, $p<$0.05; P19: $\rho$=0.23, $p<$0.05), meaning that more symptoms appeared as the average of the pupil dilation increased. We took a look at them more closely. It turned out that P13 and P19 were in the break group and also showed the lowest performance and least engagement during the study. As shown in Figure~\ref{P16_P23_and_Others}, these two participants worked primarily on low level (easy) tasks, had significantly fewer task trials, more symptoms, and lower game scores compared to the rest of the participants. This demonstrates the effect of cognitive loads and loss of concentration well but it is unclear why these two participants were affected more than others.


\section{Discussion}
 We gained a number of insights from this study. Firstly, while the total number of symptoms tended to increase over time, individual symptoms varied. Physical effects (PE), forgetting (F), inattention (I), and task rushing (TR) generally increased, whereas strategy inefficiency (STI) and task abandonment (TA) decreased over time. At the higher symptom-group level, physical and emotional deregulation (PED), and vigilance and reactionary symptoms (VR) increased, whereas personal judgement symptoms (PJS) decreased over time. This result is likely due to learning effects: as participants gained familiarity over the course of the three hour study, they became more confident in completing puzzles and had developed the most efficient strategies for doing so. But even as participants learned, cognitive depletion produced other effects to make them appear restless and impatient.
 
 The number of symptoms and that of clicks are positively correlated, which indicates that a degree of clicks could be used as a signal of symptoms. Future work will verify this result and investigate how click behavior might be used as a metric of cognitive depletion. An increase in symptoms seems to be a predictor of a break. Six participants (P1,  P3,  P9,  P10,  P15, P19) in the break group showed an above average number of symptoms at the same time as a break or one time slot before a break. This gives us an opportunity to provide systematic support (e.g., encouraging users to take a short break) to individuals when they show more symptoms over time. Particularly, some symptoms (e.g., PE, IDT, TA, INA, C) appeared more before the break compared to after breaks.
 
 Although in general both the break and non-break groups showed similar patterns of symptoms over time, TA and TUT were found to be quite different between the groups. It appears that the participants in the break group tended to be less focused than those in the non-break group. We identified additional two sub-break groups based on their performance. It turned out that the low performance group still showed an increase in PED and VR, whereas the high performance group showed a decrease. Systematic supports or suggestions could be applied to people in the low performance group to lower cognitive depletion and recover concentration.

\subsection{Limitations and Future Work}
Although our study gives us a number of interesting insights, there exist some limitations which we believe will be addressed in our future study. First, our findings may not be generalizable due to small sample size. Results could also be sensitive on other characteristics of individuals like their experience and expertise in using computer systems, concentration levels, work strategies, etc. We plan to conduct a more refined user study with more participants, and one of our focuses will be to see repeatability of the results.

Second, although we were confident that symptoms were correctly detected during the study, we also acknowledge that human errors and bias may exist. Presumably, some symptoms were easy to detect (e.g., most people would easily identify yawning) while others might be more difficult. We believe that symptoms would be better detected and understood by combining system usage logs and human annotations, which will be a primary method in our future study. 

Lastly, expanding the second point, cognitive depletion may manifest in more ways than the 14 symptoms observed in our study. For example, in many real scenarios, people sometimes work on multiple tasks. A cognitive depleted person may attempt to unnecessarily complete multiple tasks at once, increase the number of simultaneous tasks they attempt, switch between concurrent tasks more frequently, or make mid-progress switches between tasks more often \cite{dabbish2011keep,lottridge2015effects}. Moreover, a cognitively depleted person's strategy for handling pushed information (e.g., emails, updates, text messages, phone calls, etc.) will fail and become unable to continue tasks. Our current protocol and study environment was not designed to produce these conditions or provide an opportunity for related symptoms to be observed.


\section{Conclusion}
This paper presents the results of a three hour user study designed to induce cognitive depletion in 19 participants. Through the collection of different datasets including activity logs, observations, and eye-tracking data and through the individual and group-level analyses, we have gained insight into how cognitive depletion manifests in novice NMR analysts. Our work provides early insights into cognitive depletion in human computation and addresses future research implications of the findings. Future work will focus on repeatability of these results, building models that will predict cognitive status and depletion, and introducing interventions target to relieve cognitive depletion and improve overall task performance.

\bibliographystyle{aaai}
\bibliography{bibliography}

\begin{thebibliography}{}

\bibitem[\protect\citeauthoryear{Adler and Benbunan-Fich}{2013}]{adler2013self}
Adler, R.~F., and Benbunan-Fich, R.
\newblock 2013.
\newblock Self-interruptions in discretionary multitasking.
\newblock {\em Computers in Human Behavior} 29(4):1441--1449.

\bibitem[\protect\citeauthoryear{Anderson \bgroup et al\mbox.\egroup
  }{2015}]{anderson2015polymorphic}
Anderson, B.~B.; Kirwan, C.~B.; Jenkins, J.~L.; Eargle, D.; Howard, S.; and
  Vance, A.
\newblock 2015.
\newblock How polymorphic warnings reduce habituation in the brain: Insights
  from an fmri study.
\newblock In {\em Proceedings of the International SIGCHI Conference on Human
  Factors in Computing Systems},  2883--2892.
\newblock ACM.

\bibitem[\protect\citeauthoryear{Ariga and Lleras}{2011}]{ariga2011brief}
Ariga, A., and Lleras, A.
\newblock 2011.
\newblock Brief and rare mental “breaks” keep you focused: Deactivation and
  reactivation of task goals preempt vigilance decrements.
\newblock {\em Cognition} 118(3):439--443.

\bibitem[\protect\citeauthoryear{Bixler and
  D'Mello}{2013}]{bixler2013detecting}
Bixler, R., and D'Mello, S.
\newblock 2013.
\newblock Detecting boredom and engagement during writing with keystroke
  analysis, task appraisals, and stable traits.
\newblock In {\em Proceedings of the International Conference on Intelligent
  User Interfaces},  225--234.
\newblock ACM.

\bibitem[\protect\citeauthoryear{Borghini \bgroup et al\mbox.\egroup
  }{2014}]{borghini2014measuring}
Borghini, G.; Astolfi, L.; Vecchiato, G.; Mattia, D.; and Babiloni, F.
\newblock 2014.
\newblock Measuring neurophysiological signals in aircraft pilots and car
  drivers for the assessment of mental workload, fatigue and drowsiness.
\newblock {\em Neuroscience \& Biobehavioral Reviews} 44:58--75.

\bibitem[\protect\citeauthoryear{Butler \bgroup et al\mbox.\egroup
  }{2015}]{butler2015automatic}
Butler, E.; Andersen, E.; Smith, A.~M.; Gulwani, S.; and Popovi{\'c}, Z.
\newblock 2015.
\newblock Automatic game progression design through analysis of solution
  features.
\newblock In {\em Proceedings of the International SIGCHI Conference on Human
  Factors in Computing Systems},  2407--2416.
\newblock ACM.

\bibitem[\protect\citeauthoryear{Carver}{2003}]{carver2003pleasure}
Carver, C.
\newblock 2003.
\newblock Pleasure as a sign you can attend to something else: Placing positive
  feelings within a general model of affect.
\newblock {\em Cognition \& Emotion} 17(2):241--261.

\bibitem[\protect\citeauthoryear{Cheng, Teevan, and
  Bernstein}{2015}]{cheng2015measuring}
Cheng, J.; Teevan, J.; and Bernstein, M.~S.
\newblock 2015.
\newblock Measuring crowdsourcing effort with error-time curves.
\newblock In {\em Proceedings of the International SIGCHI Conference on Human
  Factors in Computing Systems},  1365--1374.
\newblock ACM.

\bibitem[\protect\citeauthoryear{Dabbish, Mark, and
  Gonz{\'a}lez}{2011}]{dabbish2011keep}
Dabbish, L.; Mark, G.; and Gonz{\'a}lez, V.~M.
\newblock 2011.
\newblock Why do i keep interrupting myself?: environment, habit and
  self-interruption.
\newblock In {\em Proceedings of the International SIGCHI Conference on Human
  Factors in Computing Systems},  3127--3130.
\newblock ACM.

\bibitem[\protect\citeauthoryear{Forster and
  Lavie}{2009}]{forster2009harnessing}
Forster, S., and Lavie, N.
\newblock 2009.
\newblock Harnessing the wandering mind: The role of perceptual load.
\newblock {\em Cognition} 111(3):345--355.

\bibitem[\protect\citeauthoryear{Galy, Cariou, and
  M{\'e}lan}{2012}]{galy2012relationship}
Galy, E.; Cariou, M.; and M{\'e}lan, C.
\newblock 2012.
\newblock What is the relationship between mental workload factors and
  cognitive load types?
\newblock {\em International Journal of Psychophysiology} 83(3):269--275.

\bibitem[\protect\citeauthoryear{Gevins and
  Smith}{2003}]{gevins2003neurophysiological}
Gevins, A., and Smith, M.~E.
\newblock 2003.
\newblock Neurophysiological measures of cognitive workload during
  human-computer interaction.
\newblock {\em Theoretical Issues in Ergonomics Science} 4(1-2):113--131.

\bibitem[\protect\citeauthoryear{Guastello \bgroup et al\mbox.\egroup
  }{2012}]{guastello2012catastrophe}
Guastello, S.~J.; Boeh, H.; Shumaker, C.; and Schimmels, M.
\newblock 2012.
\newblock Catastrophe models for cognitive workload and fatigue.
\newblock {\em Theoretical Issues in Ergonomics Science} 13(5):586--602.

\bibitem[\protect\citeauthoryear{Hart and
  Staveland}{1988}]{hart1988development}
Hart, S.~G., and Staveland, L.~E.
\newblock 1988.
\newblock Development of nasa-tlx (task load index): Results of empirical and
  theoretical research.
\newblock {\em Advances in Psychology} 52:139--183.

\bibitem[\protect\citeauthoryear{Kontogiannis and
  Malakis}{2009}]{kontogiannis2009proactive}
Kontogiannis, T., and Malakis, S.
\newblock 2009.
\newblock A proactive approach to human error detection and identification in
  aviation and air traffic control.
\newblock {\em Safety Science} 47(5):693--706.

\bibitem[\protect\citeauthoryear{Lim \bgroup et al\mbox.\egroup
  }{2010}]{lim2010imaging}
Lim, J.; Wu, W.-c.; Wang, J.; Detre, J.~A.; Dinges, D.~F.; and Rao, H.
\newblock 2010.
\newblock Imaging brain fatigue from sustained mental workload: an asl
  perfusion study of the time-on-task effect.
\newblock {\em Neuroimage} 49(4):3426--3435.

\bibitem[\protect\citeauthoryear{Lottridge \bgroup et al\mbox.\egroup
  }{2015}]{lottridge2015effects}
Lottridge, D.~M.; Rosakranse, C.; Oh, C.~S.; Westwood, S.~J.; Baldoni, K.~A.;
  Mann, A.~S.; and Nass, C.~I.
\newblock 2015.
\newblock The effects of chronic multitasking on analytical writing.
\newblock In {\em Proceedings of the 33rd International SIGCHI Conference on
  Human Factors in Computing Systems},  2967--2970.
\newblock ACM.

\bibitem[\protect\citeauthoryear{Pattyn \bgroup et al\mbox.\egroup
  }{2008}]{pattyn2008psychophysiological}
Pattyn, N.; Neyt, X.; Henderickx, D.; and Soetens, E.
\newblock 2008.
\newblock Psychophysiological investigation of vigilance decrement: boredom or
  cognitive fatigue?
\newblock {\em Physiology \& Behavior} 93(1):369--378.

\bibitem[\protect\citeauthoryear{Rzeszotarski and
  Kittur}{2011}]{rzeszotarski2011instrumenting}
Rzeszotarski, J.~M., and Kittur, A.
\newblock 2011.
\newblock Instrumenting the crowd: using implicit behavioral measures to
  predict task performance.
\newblock In {\em Proceedings of the International ACM Symposium on User
  Interface Software and Technology},  13--22.
\newblock ACM.

\bibitem[\protect\citeauthoryear{Rzeszotarski \bgroup et al\mbox.\egroup
  }{2013}]{rzeszotarski2013inserting}
Rzeszotarski, J.~M.; Chi, E.; Paritosh, P.; and Dai, P.
\newblock 2013.
\newblock Inserting micro-breaks into crowdsourcing workflows.
\newblock In {\em First AAAI Conference on Human Computation and
  Crowdsourcing}.

\bibitem[\protect\citeauthoryear{Salvucci, Taatgen, and
  Borst}{2009}]{salvucci2009toward}
Salvucci, D.~D.; Taatgen, N.~A.; and Borst, J.~P.
\newblock 2009.
\newblock Toward a unified theory of the multitasking continuum: From
  concurrent performance to task switching, interruption, and resumption.
\newblock In {\em Proceedings of the International SIGCHI Conference on Human
  Factors in Computing Systems},  1819--1828.
\newblock ACM.

\bibitem[\protect\citeauthoryear{Van~der Linden, Frese, and
  Meijman}{2003}]{van2003mental}
Van~der Linden, D.; Frese, M.; and Meijman, T.~F.
\newblock 2003.
\newblock Mental fatigue and the control of cognitive processes: effects on
  perseveration and planning.
\newblock {\em Acta Psychologica} 113(1):45--65.

\bibitem[\protect\citeauthoryear{Warm, Parasuraman, and
  Matthews}{2008}]{warm2008vigilance}
Warm, J.~S.; Parasuraman, R.; and Matthews, G.
\newblock 2008.
\newblock Vigilance requires hard mental work and is stressful.
\newblock {\em Human Factors: The Journal of the Human Factors and Ergonomics
  Society} 50(3):433--441.

\bibitem[\protect\citeauthoryear{Z{\"u}ger and
  Fritz}{2015}]{zuger2015interruptibility}
Z{\"u}ger, M., and Fritz, T.
\newblock 2015.
\newblock Interruptibility of software developers and its prediction using
  psycho-physiological sensors.
\newblock In {\em Proceedings of the Interational SIGCHI Conference on Human
  Factors in Computing Systems},  2981--2990.
\newblock ACM.

\end{thebibliography}

\end{document}